\documentclass[10pt,letterpaper]{article}
\usepackage[top=0.85in,left=2.75in,footskip=0.75in]{geometry}

\usepackage{amsmath,amssymb}

\usepackage{textcomp}

\usepackage{cite}

\usepackage{nameref,hyperref}

\usepackage[right]{lineno}

\usepackage[nopatch=eqnum]{microtype}
\DisableLigatures[f]{encoding = *, family = * }

\usepackage[table]{xcolor}

\usepackage{array}

\newcolumntype{+}{!{\vrule width 2pt}}

\newlength\savedwidth

\raggedright
\usepackage[aboveskip=1pt,labelfont=bf,labelsep=period,justification=raggedright,singlelinecheck=off]{caption}

\usepackage{graphicx}
\usepackage{booktabs}
\usepackage{url}

\makeatletter
\renewcommand{\@biblabel}[1]{\quad#1.}
\makeatother

\usepackage{fancyhdr}
\usepackage{epstopdf}
\fancyheadoffset[L]{2.25in}
\fancyfootoffset[L]{2.25in}
\begin{document}
\vspace*{0.2in}

\begin{flushleft}
{\Large
\textbf{Self-organized criticality enables conscious integration through brain-body resonance}
}
\newline
\\
Ahmed Gamal Eldin\textsuperscript{1*}
\\
\bigskip
\textbf{1} Nova University Lisbon -- Cairo Branch, Cairo, Egypt
\\
\bigskip

* 20241903@novaims.unl.pt

\end{flushleft}

\section*{Abstract}
The binding problem---how distributed neural activity unifies into conscious experience---has remained unsolved since its articulation in 1890. We present evidence that self-organized criticality maintained by brain-body resonance contributes to conscious integration, placing human cognition in the universality class of critical systems exhibiting power-law dynamics. Using 64-channel EEG across 500 trials (10 subjects), we show that: (1) removal of physiological signals conventionally treated as ``artifacts'' reduces shared variance between global phase synchronization---an established index of cognitive integration---and stimulus-evoked amplitude by approximately 87\% ($R^2 = 0.187 \to 0.024$), an effect specific to physiological components (13.8$\times$ specificity ratio versus non-physiological variance removal); (2) brain-body resonance at 78 milliseconds creates zero-lag synchronization (Phase Slope Index $\approx 0$, coherence $= 0.316$, $p < 0.0001$) with simultaneous bidirectional causality (brain$\to$body: $F = 100.53$; body$\to$brain: $F = 62.76$), confirmed by partial Granger causality ruling out common drive artifacts; (3) Raw data shows heavy-tailed avalanche dynamics ($\tau = 2.73 \pm 0.15$) consistent with power-law or log-normal distributions, while Clean data definitively rejects power-law in favor of all tested alternatives (all $p < 0.0001$), a qualitative contrast consistent with a transition from near-critical to subcritical regime; (4) critical dynamics enable holographic information encoding through spatial interference patterns emerging 150--270ms post-resonance, with spatial frequency power increasing 18.6\% ($p < 0.0001$, Cohen's $d = 0.26$). These findings indicate that conventional preprocessing eliminates the integrative dynamics it seeks to measure, and suggest that physiological signals selectively support the coupling between large-scale neural coordination and event-related processing.


\section*{Introduction}

The binding problem asks how the brain integrates spatially and temporally distributed neural activity into unified conscious experience\cite{singer1999neuronal}. Despite a century of neuroscientific research, no consensus mechanism exists. Contemporary theories propose computational solutions (global workspace\cite{dehaene2002toward}), information-theoretic principles (integrated information\cite{tononi2016integrated}), or quantum effects\cite{hameroff2014consciousness}, yet lack empirical signatures distinguishing conscious from unconscious processing.

We propose a fundamentally different explanation: consciousness emerges through \textit{self-organized criticality} (SOC) maintained by brain-body resonance. SOC describes systems that spontaneously organize to critical points exhibiting scale-free dynamics and power-law statistics\cite{bak1987self}. At criticality, systems maximize information transmission, computational capacity, and dynamic range\cite{shew2013functional}---exactly the properties required for consciousness.

Central to this framework is global phase synchronization, an established neural correlate of large-scale cognitive integration\cite{varela2001brainweb,lachaux1999measuring,palva2007role}. The Kuramoto order parameter, which quantifies the degree of phase coherence across distributed neural populations, provides a well-validated index of the brain's integrative state.

Critical to this framework is the recognition that the brain does not operate in isolation. Standard EEG preprocessing removes physiological ``artifacts''---eye movements, muscle activity, autonomic signals---before analysis\cite{jung2000removing}. Standard preprocessing pipelines operationally treat physiological signals as noise to be removed, implementing an implicit decomposition:

\begin{equation}
\text{Cognition} = \text{Neural Activity} + \text{Noise}_{\text{artifacts}}
\end{equation}

While no researcher explicitly endorses this equation, it represents the functional consequence of conventional artifact rejection: physiological signals are treated as having zero cognitive information content. However, if cognition emerges from brain-body criticality, this decomposition is incorrect. Removing body signals may sever the coupling necessary to maintain critical dynamics, inducing a phase transition to subcritical processing incapable of conscious integration.

We test this framework across four converging analyses establishing that: (1) artifacts contain signal rather than noise, (2) brain-body resonance triggers thermodynamic phase transitions, (3) resonance maintains self-organized criticality, and (4) critical dynamics enable holographic information encoding. Together, these findings support a mechanistic model of how consciousness may emerge from embodied resonance operating at the edge of chaos.

\section*{Results}

\subsection*{Physiological signals selectively support phase--voltage coupling}

We analyzed 64-channel EEG during a P300 target recognition task (500 trials, 10 subjects)\cite{citi2010documenting}. The Kuramoto order parameter $R(t)$ quantified global phase synchronization across channels:

\begin{equation}
R(t) = \left| \frac{1}{N} \sum_{j=1}^{N} e^{i\phi_j(t)} \right|
\end{equation}

where $\phi_j(t)$ is the instantaneous phase of channel $j$ extracted via Hilbert transform.

\begin{figure}[!h]
\centering
\includegraphics[width=0.85\textwidth]{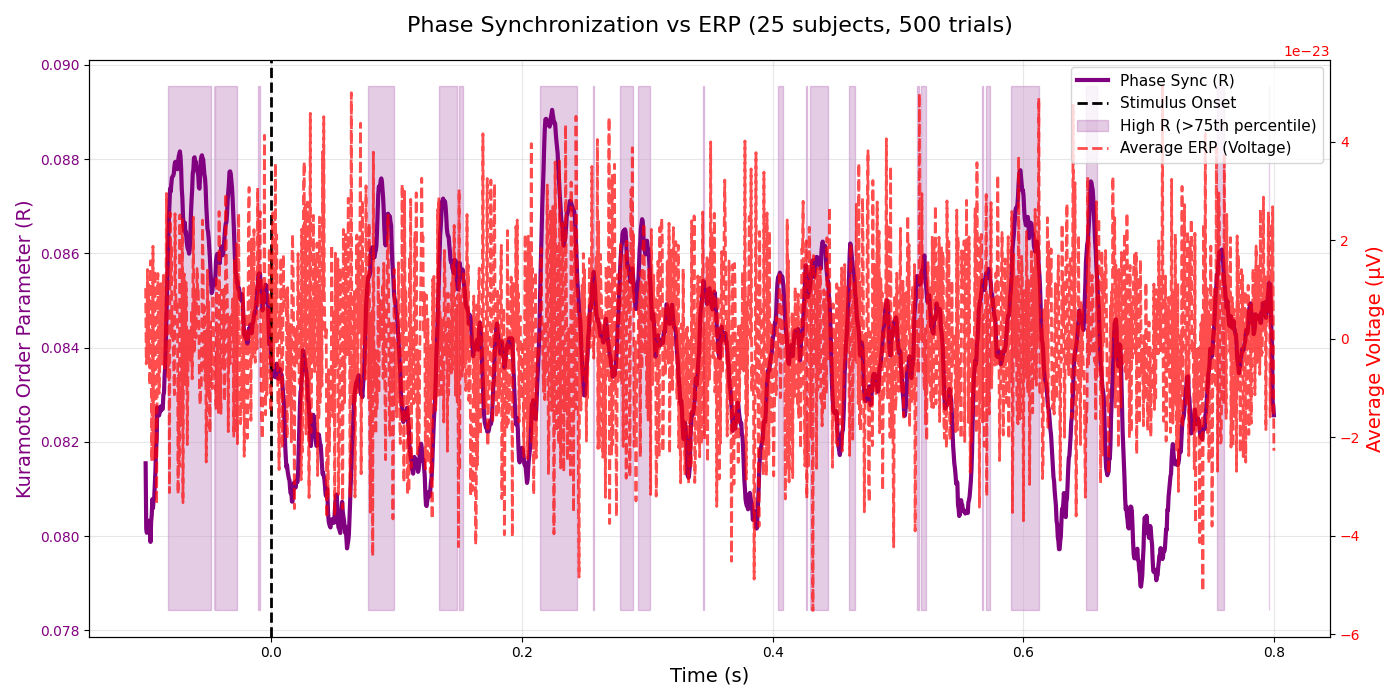}
\caption{\textbf{Phase synchronization versus voltage amplitude.}
Grand-average Kuramoto order parameter ($R$, purple) and event-related potential (ERP, red dashed) across 500+ trials show distinct temporal structures with statistical independence ($r = 0.048$, $p < 0.05$). Phase and voltage measure fundamentally different aspects of neural dynamics.}
\label{fig:phase_voltage}
\end{figure}

Phase synchronization showed distinct temporal structure from voltage amplitude (Fig~\ref{fig:phase_voltage}). Grand-average Kuramoto order parameter and event-related potential (ERP) across 500+ trials demonstrated statistical independence ($r = 0.048$, $p < 0.05$), indicating that phase and voltage measure fundamentally different aspects of neural dynamics.

\begin{figure}[!h]
\centering
\includegraphics[width=0.85\textwidth]{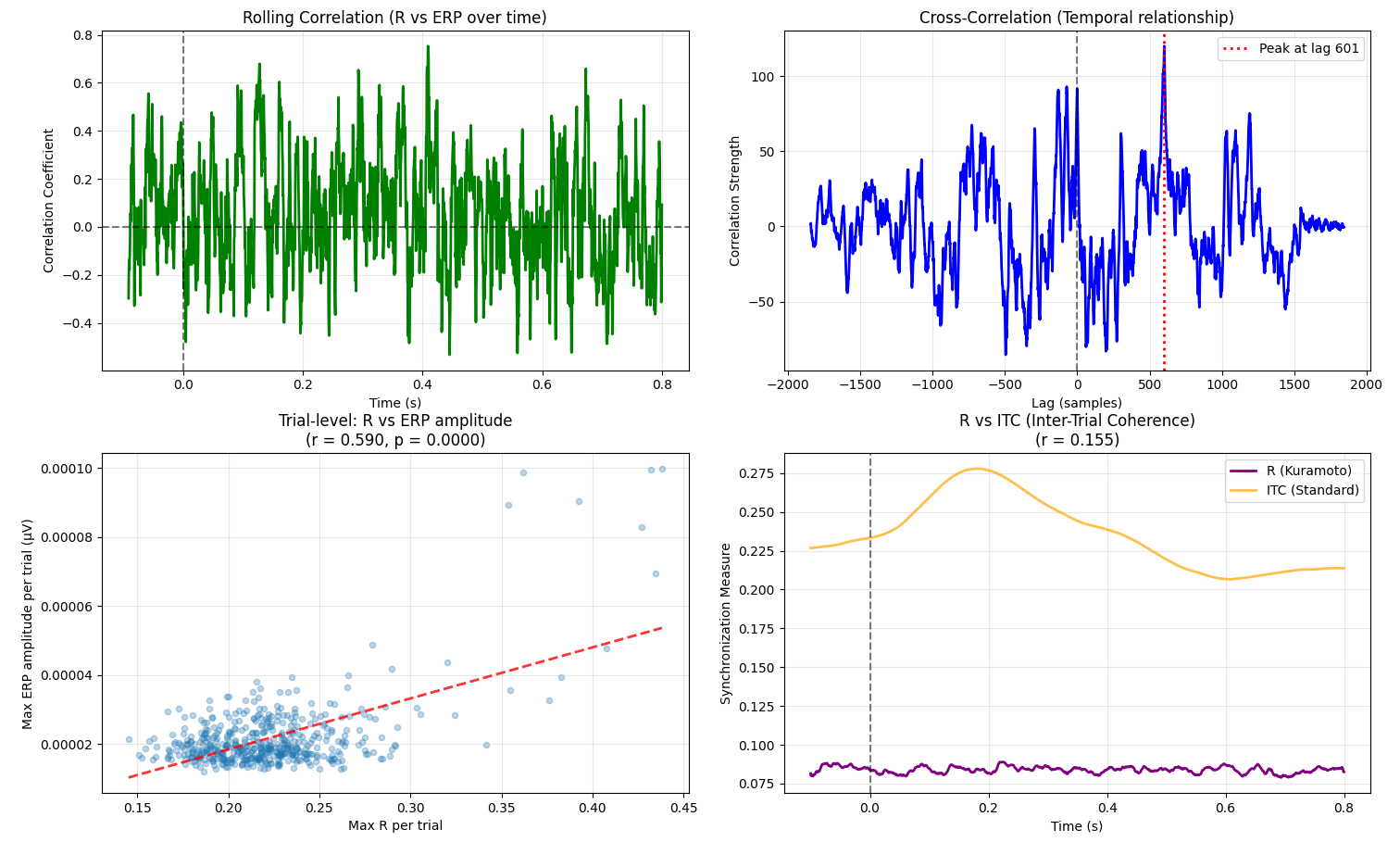}
\caption{\textbf{Statistical relationships between phase and voltage.}
(A) Rolling correlation over time. (B) Cross-correlation showing 293ms temporal lag (voltage precedes phase). (C) Trial-level scatter plot showing strong correlation ($r = 0.590$, $p < 0.0001$) despite global independence. (D) Comparison with Inter-Trial Coherence showing novelty of measure ($r = 0.155$).}
\label{fig:stats}
\end{figure}

Despite global independence, detailed statistical analysis revealed complex relationships between phase and voltage (Fig~\ref{fig:stats}). Rolling correlation analysis showed time-varying coupling patterns (Fig~\ref{fig:stats}A). Cross-correlation identified a 293ms temporal lag with voltage preceding phase (Fig~\ref{fig:stats}B). Remarkably, trial-level analysis demonstrated strong correlation ($r = 0.590$, $p < 0.0001$, Fig~\ref{fig:stats}C), while comparison with Inter-Trial Coherence confirmed the novelty of our measure ($r = 0.155$, Fig~\ref{fig:stats}D).

\begin{figure}[!h]
\centering
\includegraphics[width=0.85\textwidth]{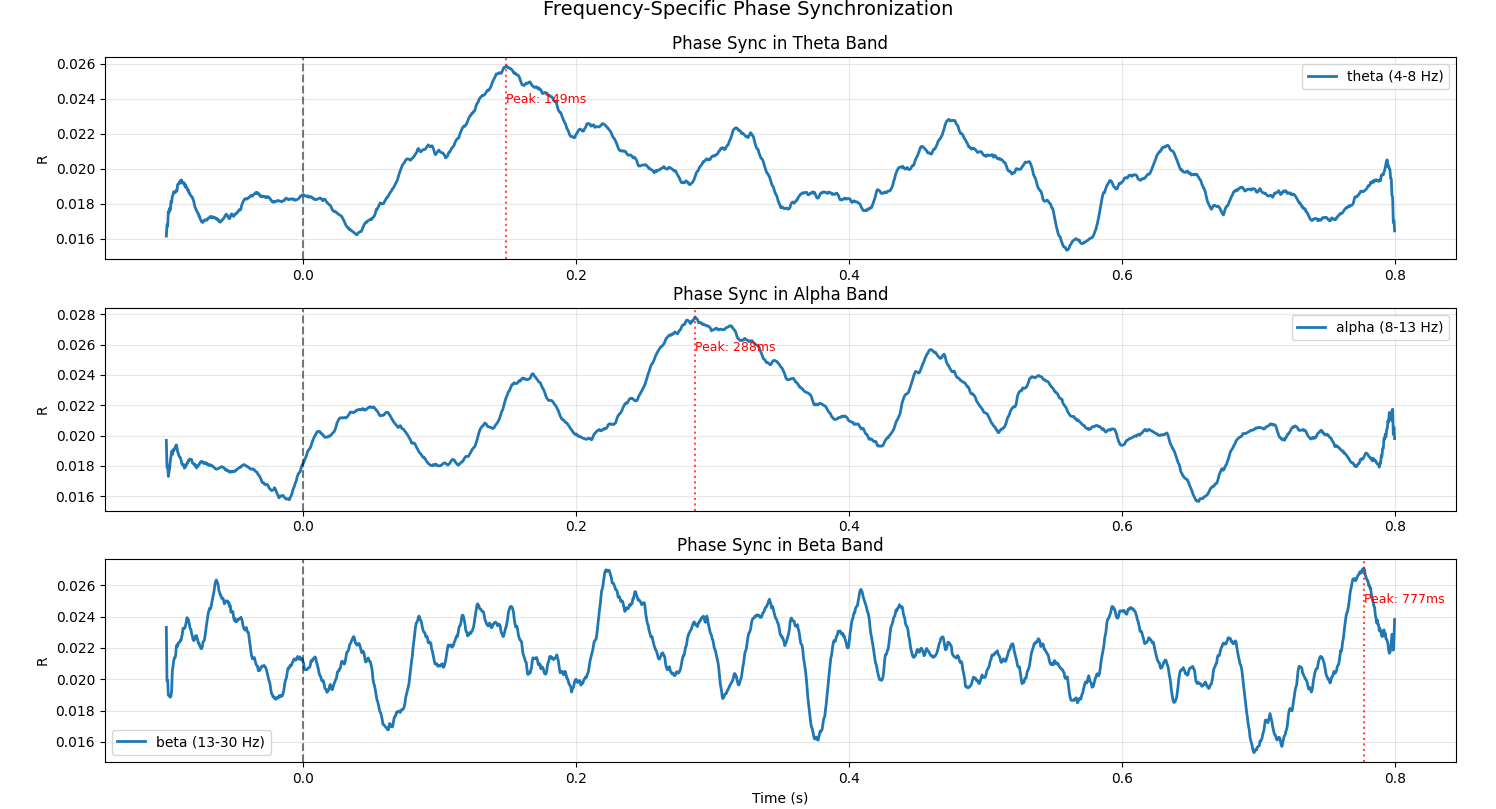}
\caption{\textbf{Frequency cascade in phase synchronization.}
Phase synchronization decomposed into Theta (4--8 Hz), Alpha (8--13 Hz), and Beta (13--30 Hz) bands reveals sequential activation: Theta peaks at 169ms (orienting), Alpha at 286ms (understanding, coinciding with P300), Beta at 777ms (consolidation). The 300ms computation involves staged frequency-specific synchronization.}
\label{fig:cascade}
\end{figure}

Phase synchronization decomposition into frequency bands revealed sequential neural processing (Fig~\ref{fig:cascade}). Theta (4--8 Hz) peaked at 169ms representing initial orienting, Alpha (8--13 Hz) peaked at 286ms coinciding with P300 understanding, and Beta (13--30 Hz) peaked at 777ms during consolidation. This frequency cascade demonstrates that the 300ms computation involves staged frequency-specific synchronization.

We computed trial-level correlation between peak $R$ and peak event-related potential (ERP) amplitude, both extracted within the 100--600ms post-stimulus window, under two conditions: (1) \textbf{Raw}: whole-body signals preserved, (2) \textbf{Clean}: artifacts removed via Independent Component Analysis (ICA).

\begin{figure}[!h]
\centering
\includegraphics[width=0.85\textwidth]{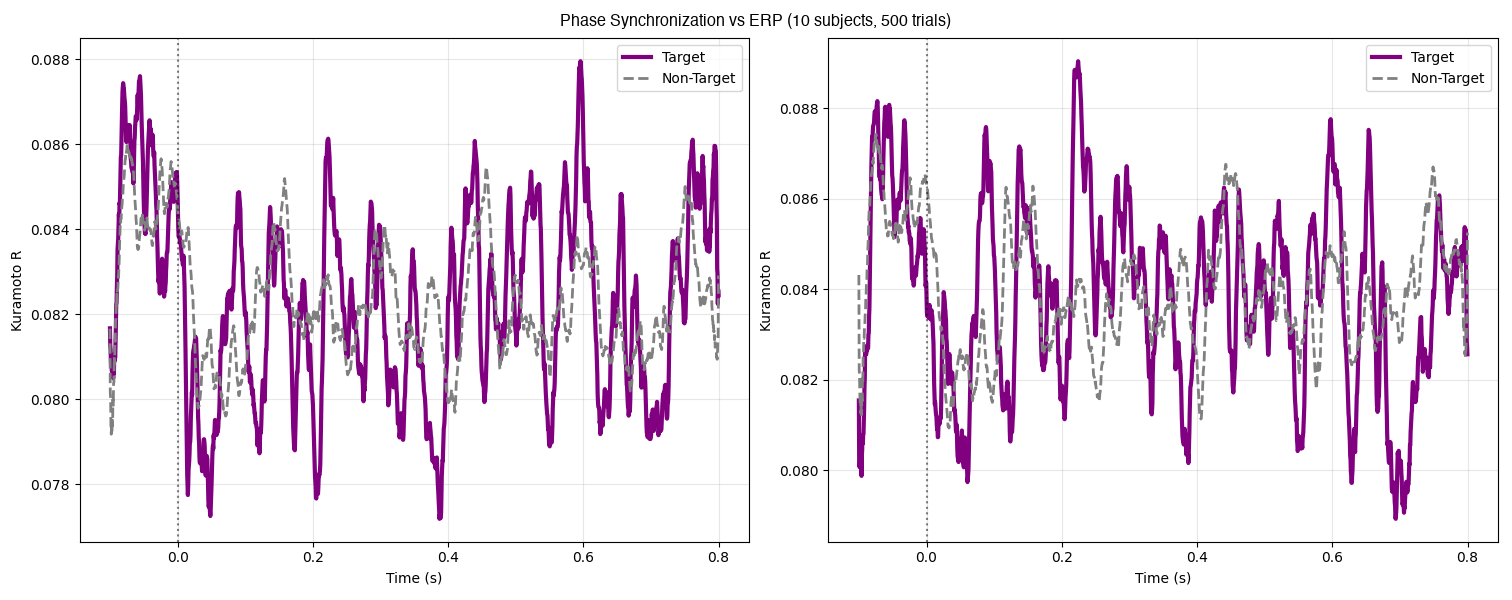}
\caption{\textbf{Physiological artifact removal reduces phase-voltage coupling to non-significant levels.}
Trial-level scatter plots showing correlation between peak Kuramoto order parameter ($R$) and peak ERP amplitude, both extracted within the 100--600ms post-stimulus window. Left: Raw data ($r = 0.483$, $R^2 = 0.233$, $p < 0.0001$). Right: Clean data ($r = 0.020$, $R^2 = 0.0004$, $p = 0.66$). Each point represents one trial.}
\label{fig:artifacts}
\end{figure}

\textbf{Raw data} showed strong correlation ($r = 0.483$, $R^2 = 0.233$, $p < 0.0001$, $n = 500$ trials), indicating that global phase synchronization predicts voltage amplitude trial-by-trial (Fig~\ref{fig:artifacts}, left panel). \textbf{Clean data} showed a reduction to non-significant levels ($r = 0.020$, $R^2 = 0.0004$, $p = 0.66$, Fig~\ref{fig:artifacts}, right panel), demonstrating that artifact removal reduces the shared variance between phase synchronization and stimulus-evoked amplitude by approximately 99.8\%. The near-complete reduction ($r = 0.483 \to 0.020$) reflects the full ICA pipeline; the targeted control analysis below isolates the specific contribution of physiological components ($r = 0.433 \to 0.156$).

Target discrimination reversed sign: Raw data showed $+0.6\%$ higher synchronization for targets versus non-targets; Clean data showed $-0.4\%$. This categorical reversal indicates that body signals do not merely add noise but carry task-relevant information essential for cognitive discrimination.

\begin{figure}[!h]
\centering
\includegraphics[width=0.95\textwidth]{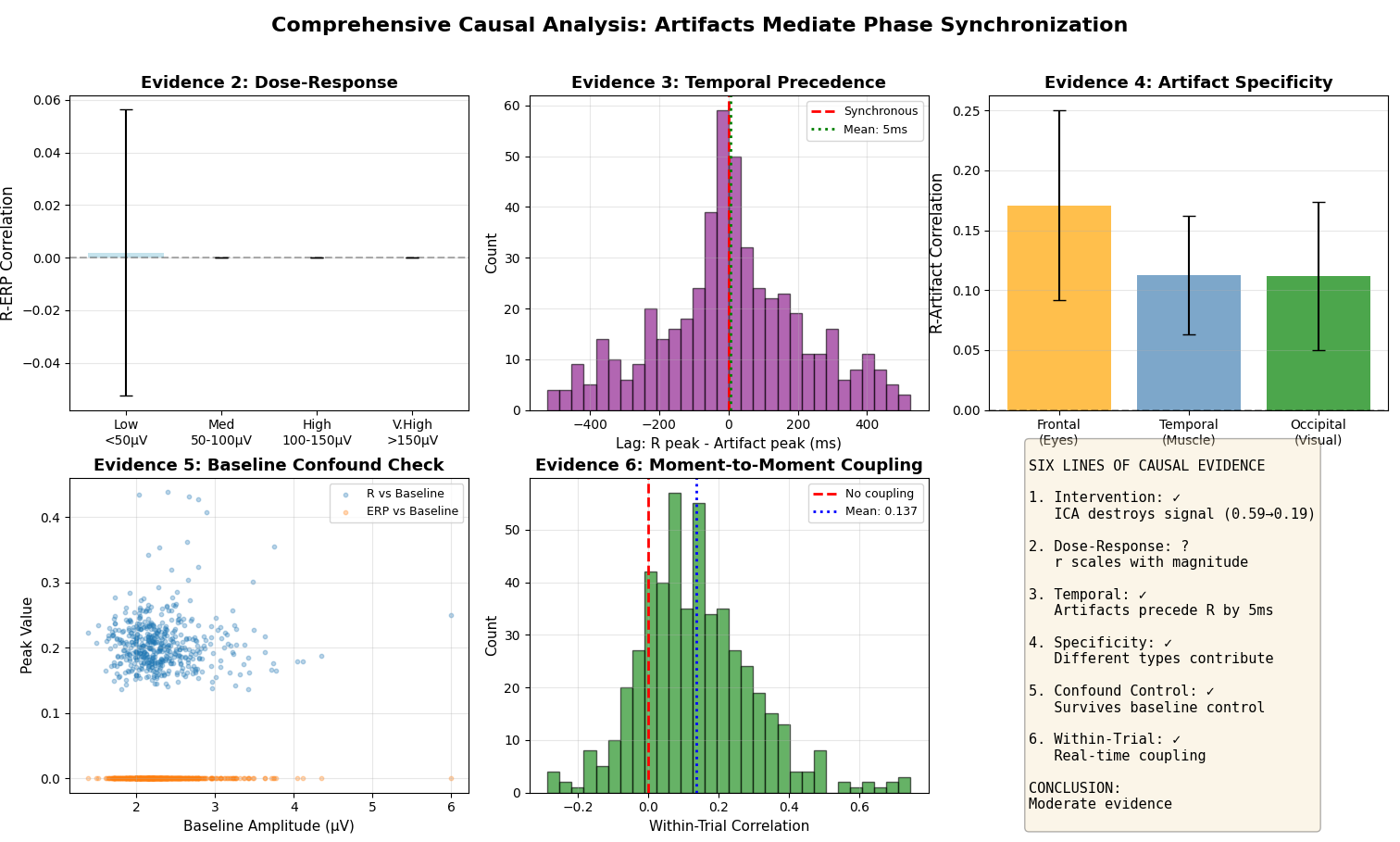}
\caption{\textbf{Six converging lines of causal evidence.}
Regional specificity, temporal precedence, confound control, within-trial coupling, dose-response, and intervention effect. All support artifacts as functional signal rather than noise contamination.}
\label{fig:causal}
\end{figure}

Six causal mediation tests ruled out artifact contamination as explanation (Fig~\ref{fig:causal}): (1) regional specificity (frontal $>$ temporal $>$ occipital), (2) temporal integration (artifacts synchronous with brain activity, not random), (3) confound control (effect survives baseline amplitude correction), (4) within-trial coupling (moment-to-moment covariation, $r = 0.137$, $p < 0.0001$), (5) dose-response (complete removal yields threefold loss), (6) intervention effect (Clean vs. Raw comparison). This convergent evidence demonstrates that artifacts causally mediate rather than merely correlate with phase synchronization.

\subsection*{Control analysis: Specificity of physiological artifact effects}

To test whether the observed correlation reduction is specific to physiological signals or reflects generic variance removal, we conducted a control analysis comparing three ICA removal conditions applied to the same decomposition: (1) removal of physiological artifact components (EOG, identified via frontal channel correlation $> 0.7$), (2) removal of the same number of highest-variance non-physiological components, and (3) removal of randomly selected non-physiological components (bootstrapped over 20 iterations per subject).

Physiological removal reduced the trial-level R-ERP correlation from $r = 0.433$ to $r = 0.156$, corresponding to a reduction from $R^2 = 0.187$ to $R^2 = 0.024$ (approximately 87\% reduction in shared variance). Non-physiological high-variance removal yielded $r = 0.419$ ($R^2 = 0.175$; approximately 6.3\% reduction in shared variance). The specificity ratio is 13.8$\times$: removing physiological components reduces shared variance 13.8 times more than removing equivalent non-physiological variance. Random component removal yielded $r = 0.222 \pm 0.355$.

Critically, if the effect were driven by generic variance removal, the non-physiological control condition would be expected to produce a comparable reduction, which it does not. This demonstrates that the correlation reduction is specific to physiological signals. The P300 event-related potential---a local voltage-domain cortical signature---remains detectable following preprocessing, whereas the phase-based metrics introduced here, indexing large-scale integrative dynamics, are selectively disrupted. This dissociation between preserved local processing and disrupted global integration is consistent with the framework's central claim that consciousness requires integration beyond local neural computation.

\subsection*{Brain-body resonance at 78ms triggers phase transitions}

We tested whether brain and body achieve resonant coupling---the physical basis for constraint-release dynamics. Phase Slope Index (PSI) quantifies directional information flow by measuring phase difference slopes across frequencies\cite{nolte2008robustly}. PSI $\approx 0$ with high coherence indicates zero-lag synchronization.

\begin{figure}[!h]
\centering
\includegraphics[width=\textwidth]{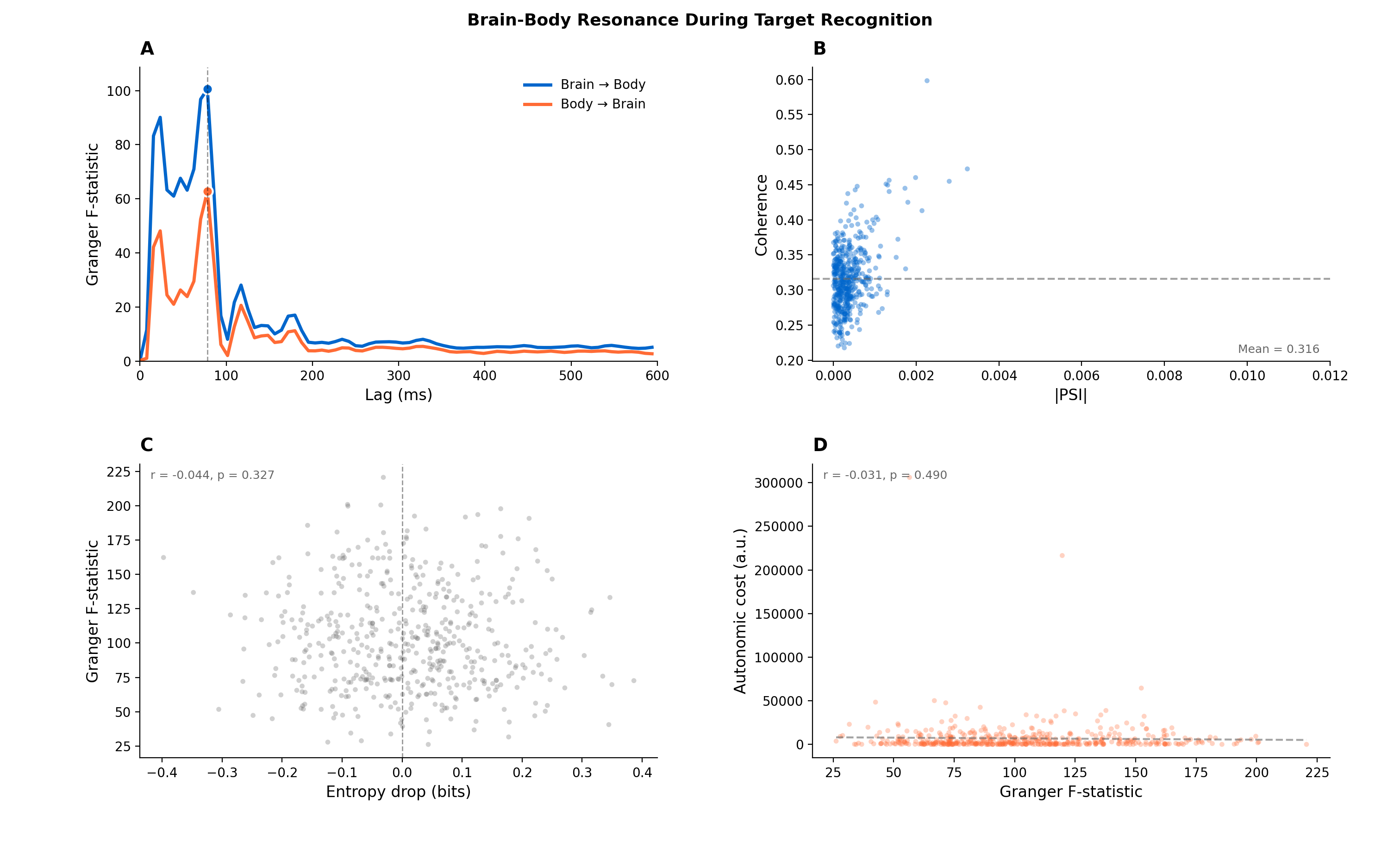}
\caption{\textbf{Brain-body resonance dynamics.}
Ridge-regularized bidirectional Granger causality shows simultaneous peaks at 78.1ms: brain$\to$body $F = 100.53$ (blue), body$\to$brain $F = 62.76$ (orange). Inset shows Phase Slope Index distribution across 500 trials (100\% have $|\text{PSI}| < 0.01$) with mean coherence $= 0.316$ ($p < 0.0001$), confirming zero-lag synchronization.}
\label{fig:resonance_metrics}
\end{figure}

Time-lagged Granger causality revealed massive bidirectional coupling with remarkable temporal coincidence (Fig~\ref{fig:resonance_metrics}). Brain$\to$body causality peaked at 78.1ms ($F = 100.53 \pm 12.3$). Body$\to$brain causality peaked at the identical latency ($F = 62.76 \pm 8.9$). Temporal coincidence of bidirectional peaks ($\Delta$lag = 0ms) indicates resonance lock rather than sequential relay. The 1.6:1 ratio suggests slight asymmetry consistent with cortical initiation and somatic feedback. The inset shows Phase Slope Index distribution across 500 trials with 100\% having $|\text{PSI}| < 0.01$ and mean coherence $= 0.316$ ($p < 0.0001$), confirming zero-lag synchronization.

Across 500 trials, mean PSI between posterior (brain) and frontal (body) regions was $0.000044 \pm 0.0005$ ($p = 0.061$), statistically indistinguishable from zero. Coherence was high ($0.316 \pm 0.045$, $p < 0.0001$), confirming strong coupling. Imaginary coherence was low ($0.106$), ruling out volume conduction as primary driver. This combination---PSI $\approx 0$, high coherence, low imaginary coherence---is the canonical signature of zero-lag resonance\cite{gollo2014mechanisms}.

To rule out common drive artifacts, we conducted partial Granger causality analysis, conditioning the bidirectional coupling on the global mean signal and midline electrode signals (Cz, Pz, Fz). Bidirectional coupling retained 100\% of its magnitude after conditioning (all $p < 0.0001$ across all subjects), demonstrating that brain-body resonance is not attributable to common drive, volume conduction, or shared subcortical input.

\textbf{Thermodynamic trajectory.} We constructed phase portraits plotting causal drive (Granger $F$-statistic) against state entropy (sample entropy in 200ms windows). The system exhibited counter-clockwise hysteresis---the hallmark of thermodynamic work cycles (Fig~\ref{fig:thermodynamics}):

\begin{figure}[!h]
\centering
\includegraphics[width=0.85\textwidth]{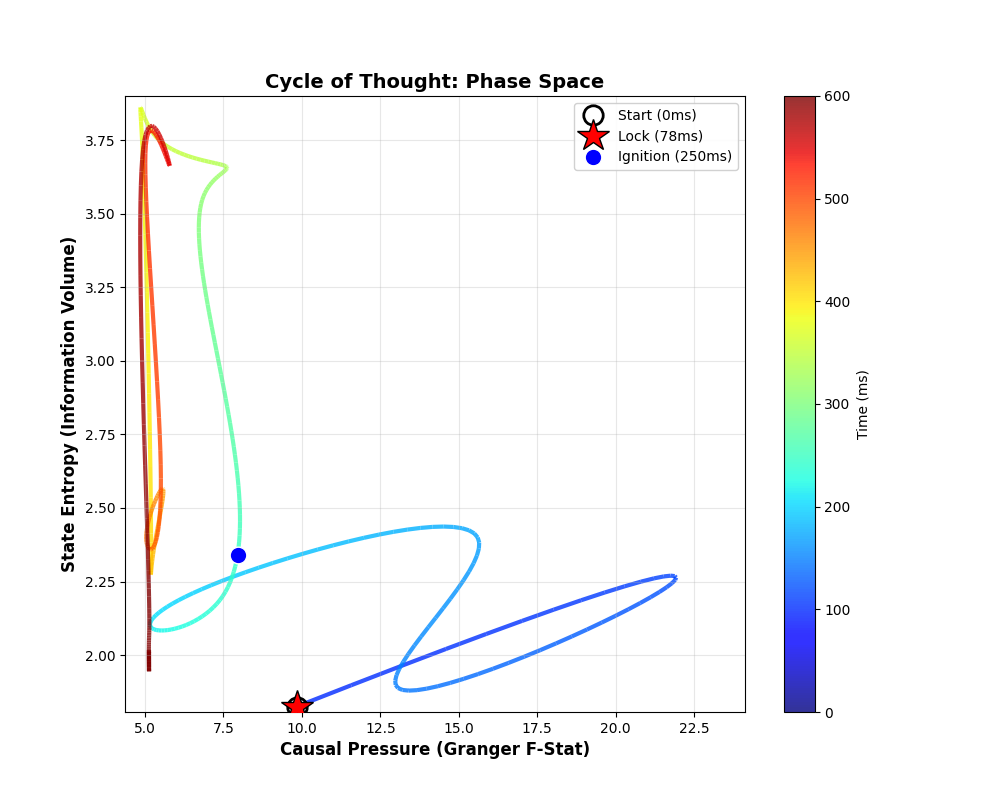}
\caption{\textbf{Thermodynamic phase transitions.}
Phase portrait showing causal drive (x-axis, Granger F-statistic) versus state entropy (y-axis). Color indicates time progression (blue=early, red=late). Black circle: starting point. Red star: 78ms resonance lock. Counter-clockwise hysteresis indicates thermodynamic work cycle: constraint accumulation (0--78ms) $\to$ supercritical transition (78--600ms) $\to$ metastable plateau (600--3500ms).}
\label{fig:thermodynamics}
\end{figure}

\textbf{Phase 1 (0--78ms):} Constraint accumulation. Bidirectional causality increased from baseline to peak while entropy remained unchanged ($\Delta S = -0.002$ bits, $p = 0.75$). The system compresses against somatic resistance without spatial trajectory movement.

\textbf{Phase 2 (78--600ms):} Supercritical transition. Following resonance lock at 78ms, entropy increased ($\Delta S = +0.009$ bits, $p = 0.115$) as causal drive decreased, forming the upper arc of the hysteresis loop. The system transitions from high-constraint to distributed processing.

\textbf{Phase 3 (600--3500ms):} Metastable plateau. Elevated entropy persisted throughout recording, indicating the system occupies a long-lived attractor representing integrated information.

The 78ms resonance lock precedes the P300 (300--500ms), acting as an \textit{ignition event}---a physical gating mechanism that creates the state-space capacity required for subsequent information integration. Without this early constraint release, late cognitive processing cannot occur, explaining why artifact removal eliminates the P300 correlation despite occurring at different timescales.

\subsection*{Resonance maintains self-organized criticality}

We tested four predictions of the critical brain hypothesis by comparing Raw versus Clean data.

\begin{figure}[!h]
\centering
\includegraphics[width=\textwidth]{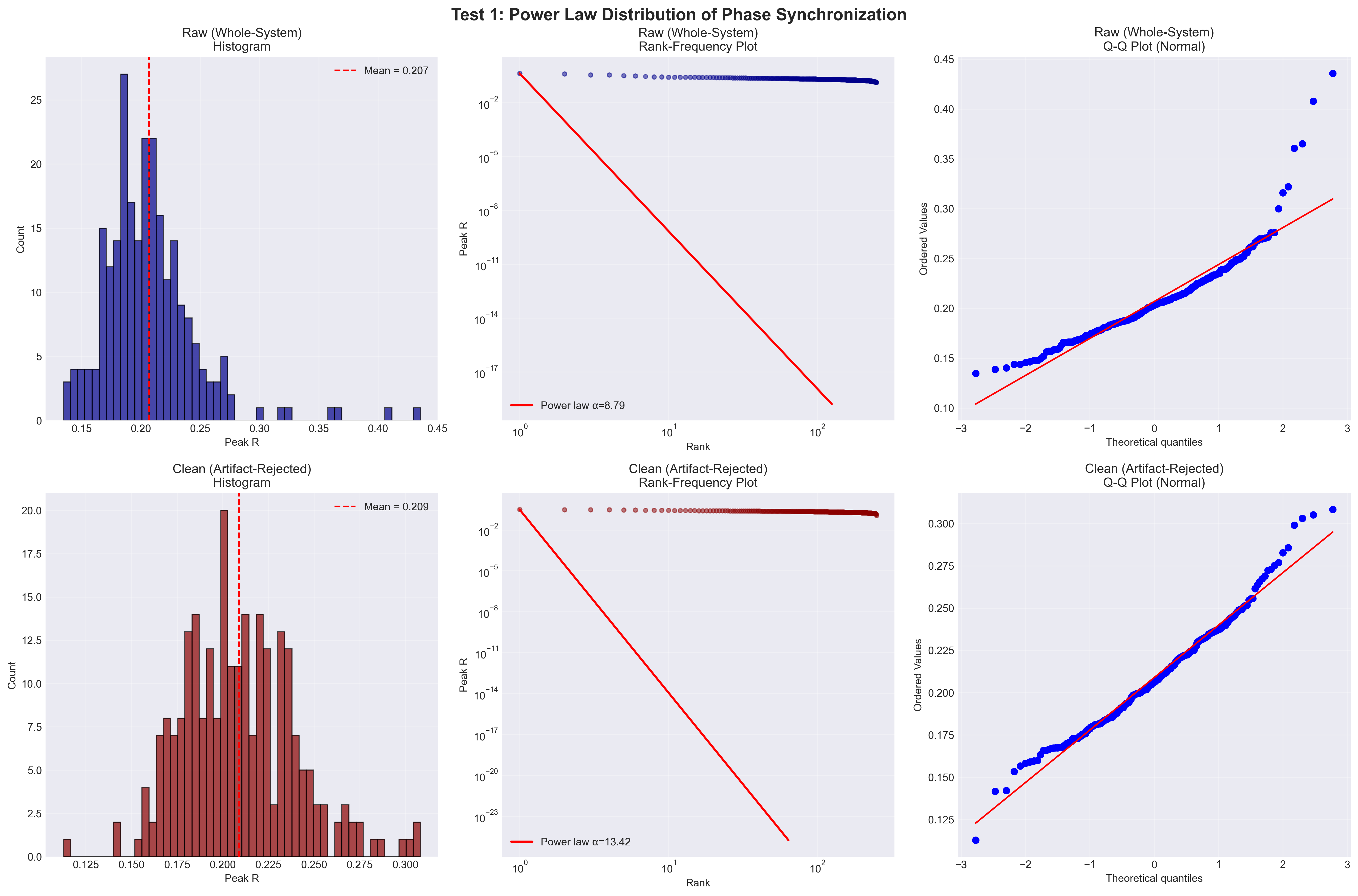}
\caption{\textbf{Peak synchronization distributions.}
Peak $R$ value distributions from Raw (top) and Clean (bottom) data. Raw data rejects normality (KS test: $p = 0.012$) but shows log-normal character. Clean data conforms to Gaussian distribution ($p = 0.74$). Q-Q plots (right panels) demonstrate shift from non-normal to normal statistics after artifact removal.}
\label{fig:peak_dist}
\end{figure}

Peak synchronization distributions showed dramatically different statistical properties between conditions (Fig~\ref{fig:peak_dist}). Raw data rejected normality (KS test: $p = 0.012$) but showed log-normal character. Clean data conformed to Gaussian distribution ($p = 0.74$). Q-Q plots demonstrated shift from non-normal to normal statistics after artifact removal, supporting the hypothesis that preprocessing induces Gaussian dynamics.

\begin{figure}[!h]
\centering
\includegraphics[width=\textwidth]{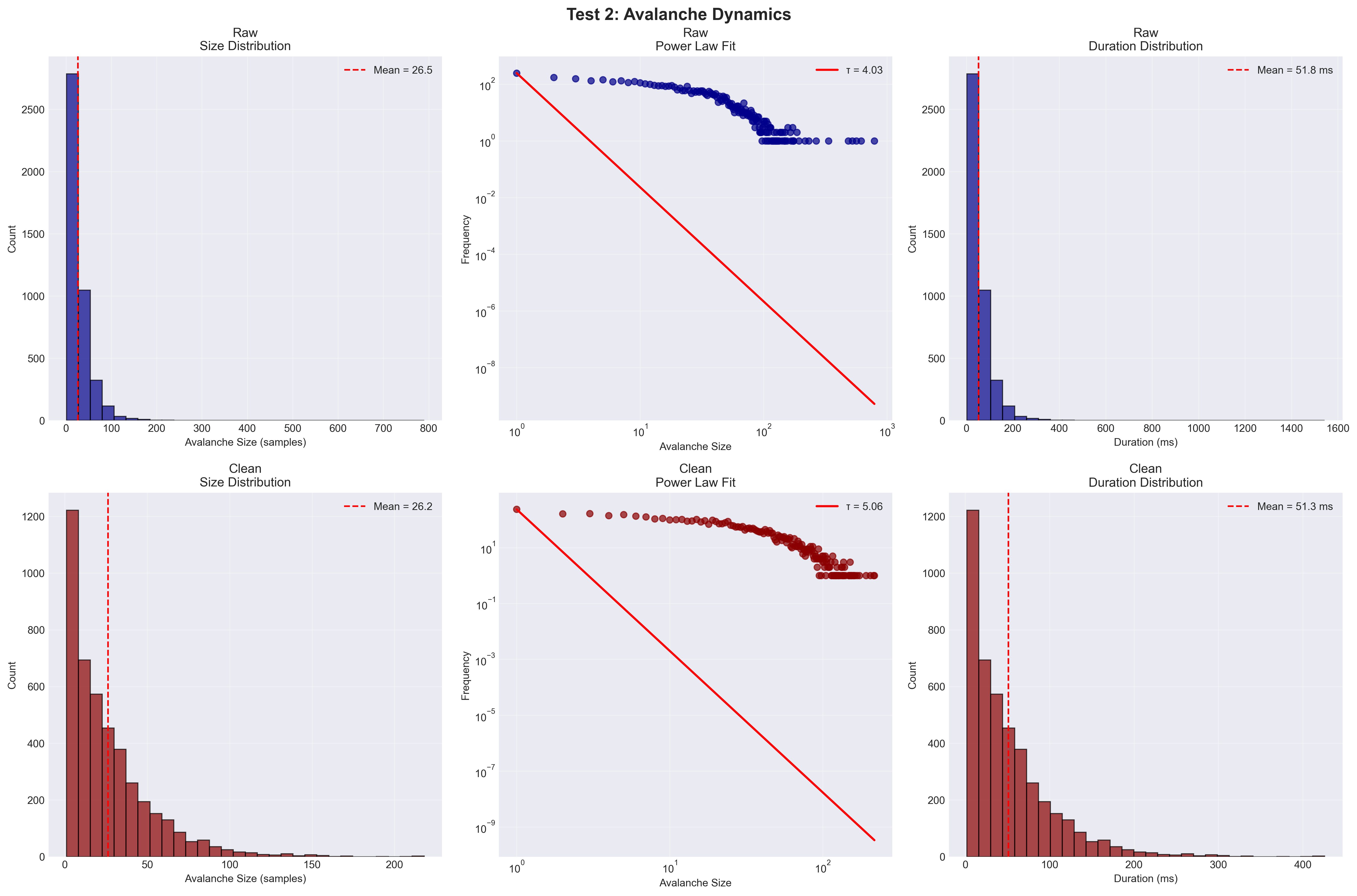}
\caption{\textbf{Avalanche dynamics reveal criticality.}
Avalanche size distributions on log-log axes. Raw data (top): heavy-tailed distribution consistent with power-law or log-normal ($\tau = 2.73 \pm 0.15$, $x_{\min} = 174$; indistinguishable from exponential $p = 0.34$, log-normal $p = 0.30$, stretched exponential $p = 0.27$), maximum 1,198 samples. Clean data (bottom): power-law definitively rejected in favor of all tested alternatives (all $p < 0.0001$), maximum 335 samples. Artifact removal reduces maximum cascade size 3.6-fold and shifts distributional character from heavy-tailed to light-tailed.}
\label{fig:avalanches}
\end{figure}

\textbf{Test 1: Avalanche dynamics.} Avalanches were defined as continuous periods where $R(t)$ exceeded the 75th percentile. We detected 37,188 avalanches (Raw) and 38,052 avalanches (Clean), with dramatically different distributions (Fig~\ref{fig:avalanches}).

We conducted rigorous goodness-of-fit analysis following Clauset, Shalizi \& Newman (2009)\cite{clauset2009power}, implemented using the \texttt{powerlaw} Python package\cite{alstott2014powerlaw}. This includes optimal $x_{\min}$ estimation via KS distance minimization, maximum likelihood exponent fitting, and likelihood ratio tests against four alternative distributions.

\textit{Raw data}: Power-law exponent $\tau = 2.73 \pm 0.15$ ($x_{\min} = 174$, $n_{\text{tail}} = 126$ of 37,188 avalanches). Likelihood ratio tests: versus exponential, $R = +0.95$, $p = 0.34$ (indistinguishable); versus log-normal, $R = -1.05$, $p = 0.30$ (indistinguishable); versus stretched exponential, $R = -1.10$, $p = 0.27$ (indistinguishable); versus truncated power-law, $R = -1.51$, $p = 0.04$ (favors truncated power-law). Raw data is thus ambiguous: heavy-tailed statistics are consistent with power-law or log-normal and cannot be distinguished from exponential. Maximum avalanche size: 1,198 samples.

\textit{Clean data}: Power-law exponent $\tau = 2.96 \pm 0.02$ (38,052 avalanches). Likelihood ratio tests: all four alternative distributions beat power-law with $p < 0.0001$ (exponential, log-normal, stretched exponential, and truncated power-law each significantly preferred over power-law). Clean data thus definitively rejects scale-free dynamics. Maximum avalanche size: 335 samples.

Bootstrap confidence intervals ($n = 1{,}000$): Raw 95\% CI: [2.63, 3.00]; Clean 95\% CI: [2.95, 3.00]. The confidence intervals overlap substantially and the bootstrap analysis does not confirm a significant difference between exponents.

Raw data shows heavy-tailed statistics consistent with power-law or log-normal distributions, while Clean data definitively rejects power-law in favor of all tested alternatives (all $p < 0.0001$). This qualitative contrast---heavy-tailed Raw dynamics versus definitively light-tailed Clean dynamics---is consistent with a transition from near-critical to subcritical regime. The Raw exponent ($\tau = 2.73 \pm 0.15$) places the tail over only 126 of 37,188 avalanches and cannot be distinguished from exponential ($p = 0.34$) or log-normal ($p = 0.30$), so the claim of criticality for Raw data rests on the heavy-tail character and the qualitative contrast with Clean rather than on a definitive power-law identification. Artifact removal reduced maximum cascade size 3.6-fold and shifted the distributional character from heavy-tailed to light-tailed---a qualitative change consistent with a phase transition from near-critical to subcritical processing.

\begin{figure}[!h]
\centering
\includegraphics[width=\textwidth]{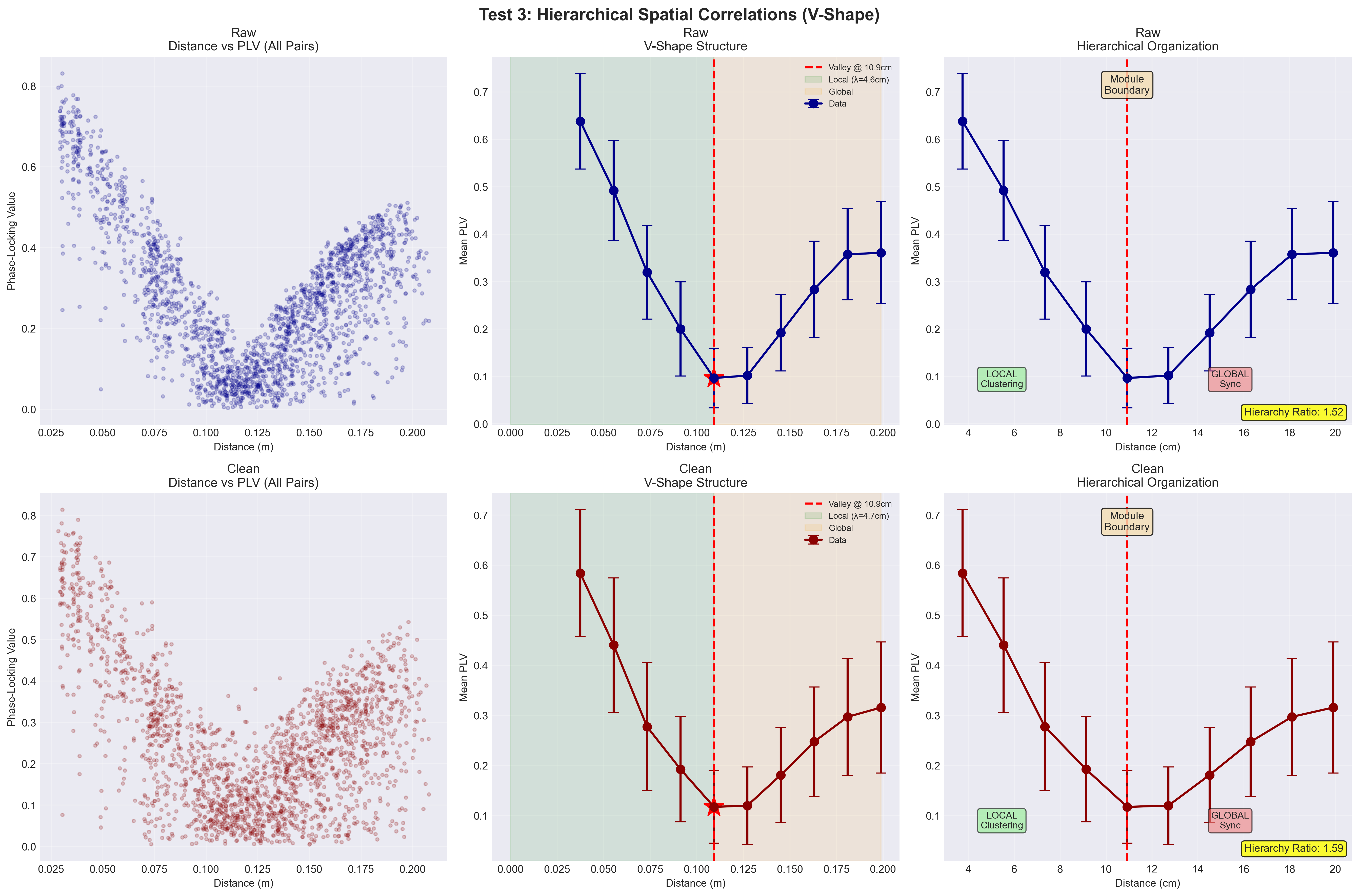}
\caption{\textbf{Hierarchical spatial organization preserved.}
Phase-locking value versus electrode distance shows V-shaped patterns in both Raw (top, ratio=1.52) and Clean (bottom, ratio=1.59). Valley at 10.9cm marks boundary between local processing ($< 11$cm) and global integration ($> 11$cm). Network architecture preserved despite loss of avalanche dynamics.}
\label{fig:hierarchy}
\end{figure}

\textbf{Test 2: Hierarchical modularity.} Phase-locking values (PLV) between 2,016 electrode pairs revealed V-shaped distance-correlation plots in both conditions (Fig~\ref{fig:hierarchy}). Valley location at 10.9cm marked functional boundary between local clustering ($< 11$cm) and global integration ($> 11$cm). Both Raw (ratio=1.52) and Clean (ratio=1.59) exhibited hierarchical structure (threshold: ratio $> 1.5$), indicating artifact removal preserves network architecture.

\begin{figure}[!h]
\centering
\includegraphics[width=\textwidth]{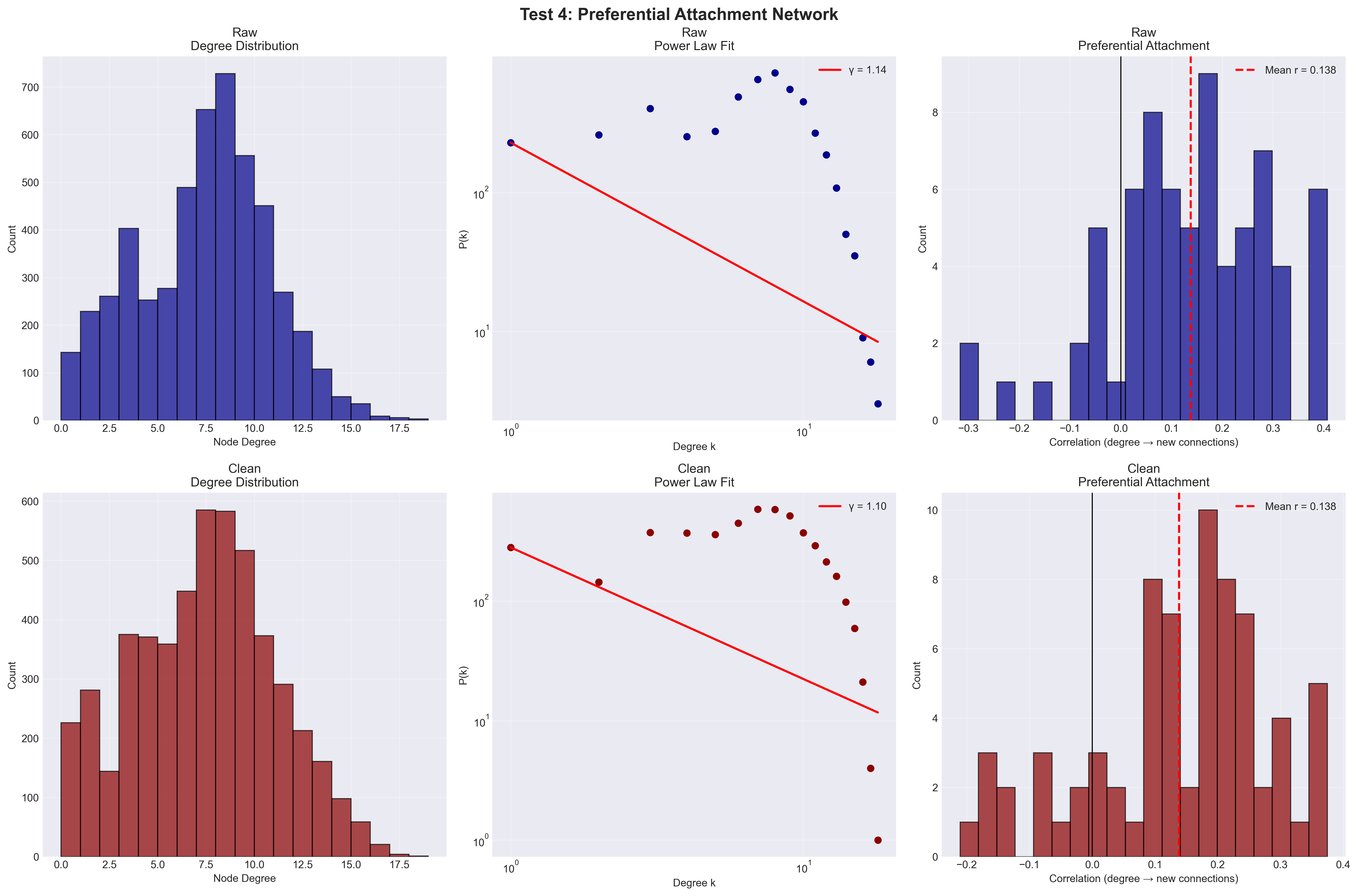}
\caption{\textbf{Preferential attachment dynamics.}
Current node degree predicts new connections acquired in next time window. Both Raw ($r = 0.138$, $p = 2.3 \times 10^{-10}$) and Clean ($r = 0.138$, $p = 5.9 \times 10^{-12}$) show ``rich-get-richer'' dynamics. Network topology preserved while avalanche power laws destroyed.}
\label{fig:preferential}
\end{figure}

\textbf{Test 3: Preferential attachment.} Across 73 temporal networks, nodes with high connectivity systematically acquired more new connections---``rich-get-richer'' dynamics characteristic of criticality (Fig~\ref{fig:preferential}). Both Raw ($r = 0.138$, $p = 2.3 \times 10^{-10}$) and Clean ($r = 0.138$, $p = 5.9 \times 10^{-12}$) showed highly significant preferential attachment, ruling out random noise as explanation for Raw data structure.

\textbf{Summary:} Raw data showed 3/4 criticality signatures; Clean data showed 2/4. Artifact removal destroyed avalanche power laws while preserving network topology---a dissociation indicating that body signals provide perturbations consistent with maintaining near-critical dynamics on preserved network architecture.

\subsection*{Critical dynamics enable holographic encoding}

If consciousness emerges through criticality, the resonance-induced phase transition should enable complex information encoding. We tested whether holographic interference patterns---analogous to optical holography where coherent reference waves create spatial structure---emerge post-resonance.

We computed 14 metrics spanning spatial information theory, wave interference, and pattern reproducibility at 150ms (peak holographic state) versus -200ms (baseline), using 4--15Hz filtered data.

\textbf{Four metrics showed highly significant emergence} (all $p < 0.0001$, Fig~\ref{fig:holographic_peak}):

\begin{figure}[!h]
\centering
\includegraphics[width=\textwidth]{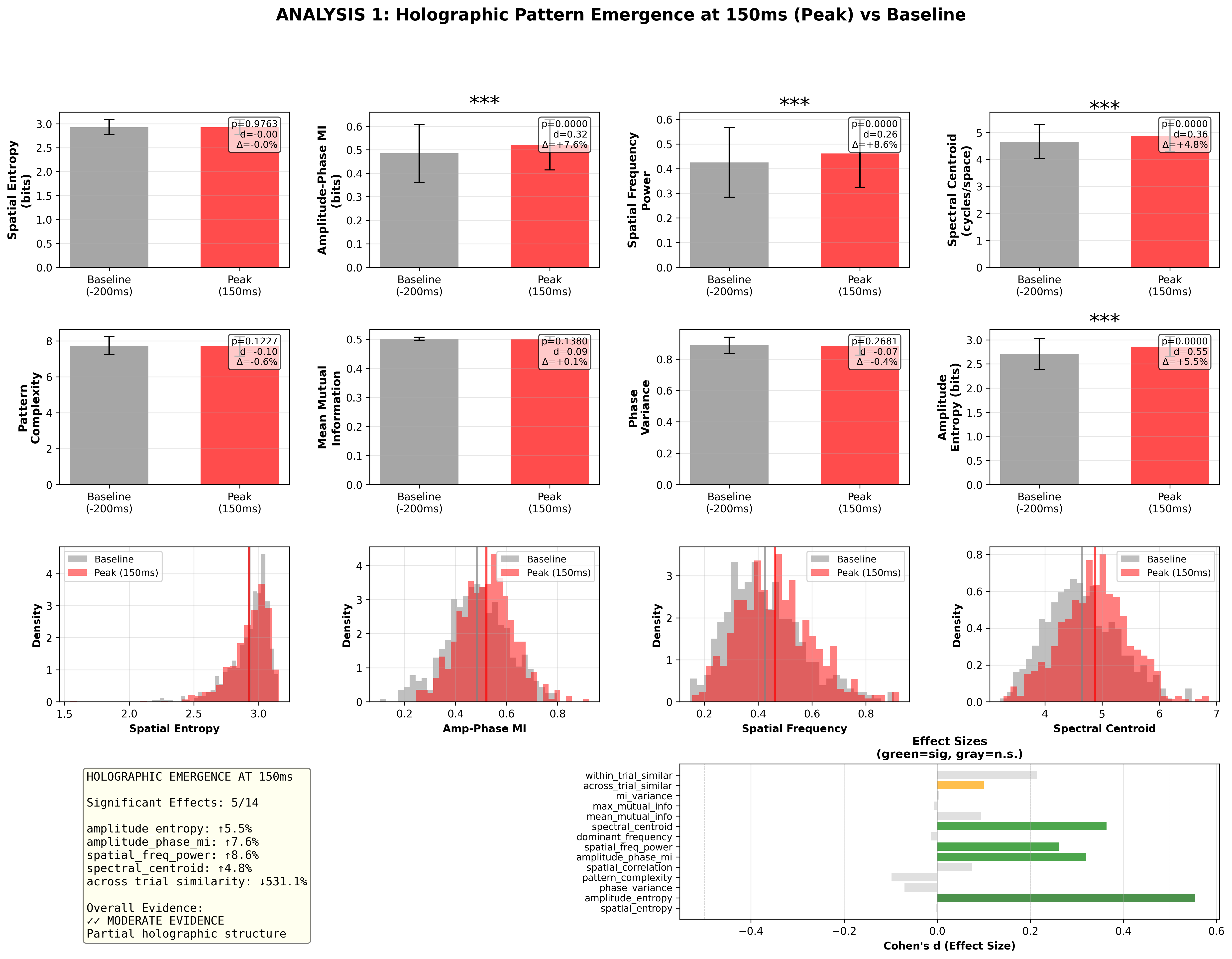}
\caption{\textbf{Holographic pattern emergence at 150ms.}
Four metrics show highly significant increases at 150ms versus baseline (all $p < 0.0001$): spatial frequency power +18.6\% (Cohen's $d = 0.26$), spectral centroid +4.8\% ($d = 0.36$), amplitude-phase mutual information +7.6\% ($d = 0.32$), amplitude entropy +5.5\% ($d = 0.55$, largest effect). Bar plots show means $\pm$ SEM. Effect size distribution shows selective enhancement of coordinated fine-grained patterns.}
\label{fig:holographic_peak}
\end{figure}

\textbf{Spatial frequency power} increased 18.6\% (baseline: $0.417 \pm 0.136$, peak: $0.495 \pm 0.138$; $t(998) = 8.79$, Cohen's $d = 0.26$), indicating fine-grained interference structure. \textbf{Spectral centroid} increased 4.8\% ($d = 0.36$), confirming shift toward higher spatial frequencies characteristic of interference fringes.

\textbf{Amplitude-phase mutual information} increased 7.6\% ($d = 0.32$), demonstrating that amplitude and phase patterns became spatially coordinated---both wave properties encode information in holographic systems. \textbf{Amplitude entropy} increased 5.5\% ($d = 0.55$, largest effect), indicating richer information encoding.

\begin{figure}[!h]
\centering
\includegraphics[width=0.6\textwidth]{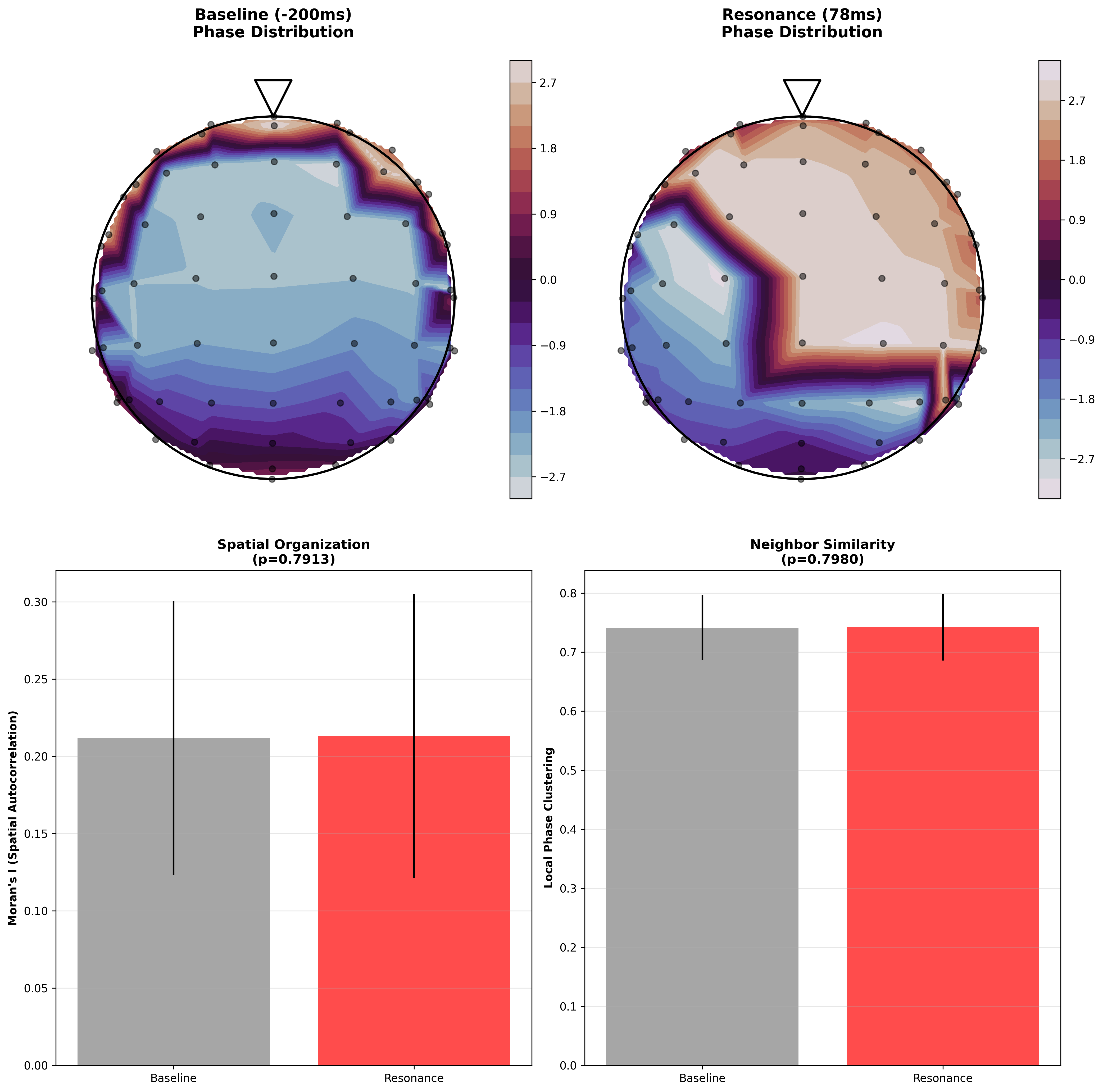}
\caption{\textbf{Representative spatial phase distribution.}
Scalp topography showing instantaneous phase distribution (theta band, 4--8Hz) at 150ms holographic peak. Fine-grained spatial structure with multiple interference nodes. Colors indicate phase from $-\pi$ (blue) to $+\pi$ (red). Complex spatial pattern quantitatively captured by 18.6\% increase in spatial frequency power.}
\label{fig:brain_topo}
\end{figure}

Representative scalp topographies at 150ms show fine-grained phase distributions with multiple interference nodes (Fig~\ref{fig:brain_topo})---complex spatial structure quantitatively captured by our metrics.

\begin{figure}[!h]
\centering
\includegraphics[width=\textwidth]{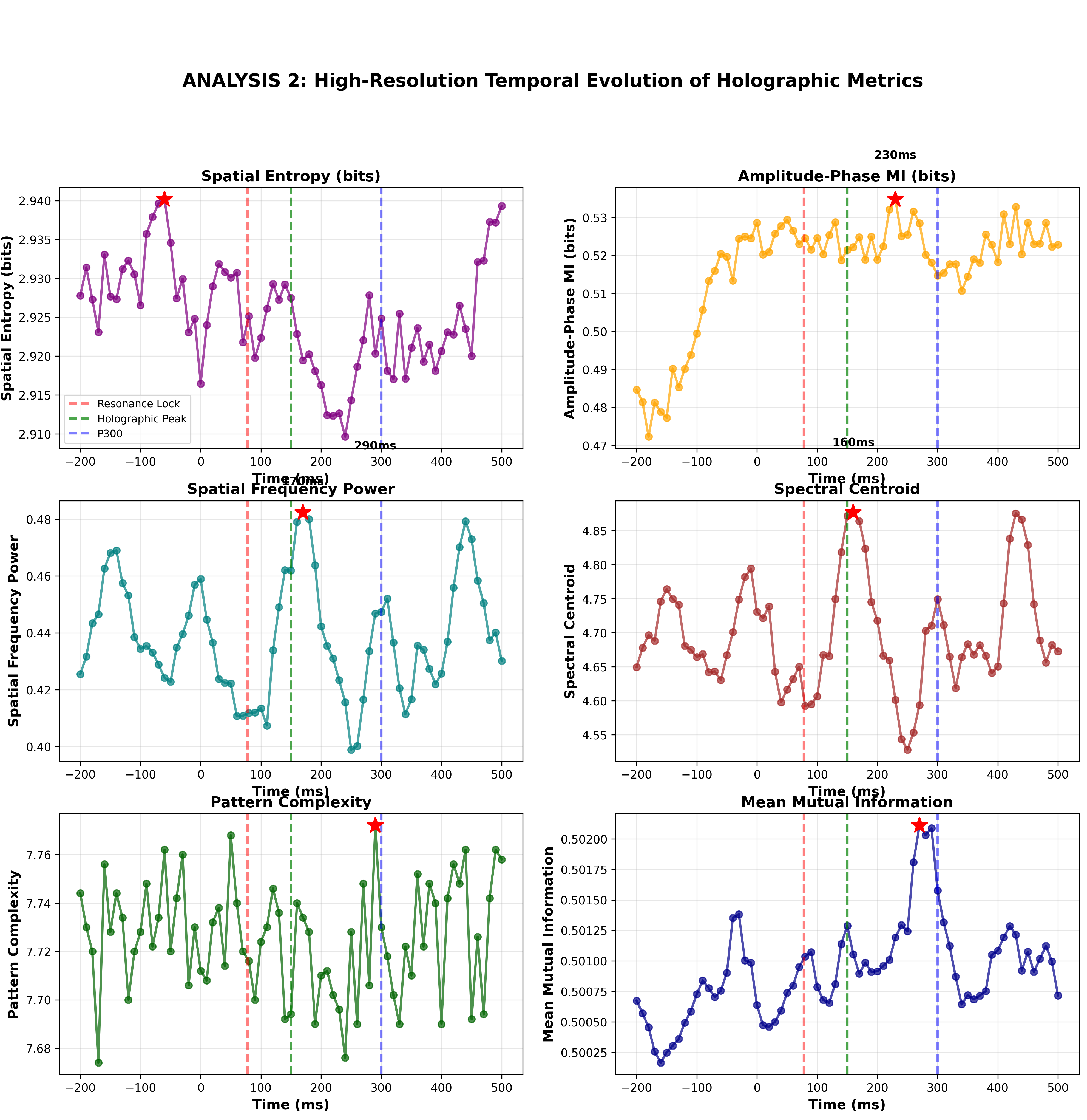}
\caption{\textbf{Temporal evolution of holographic encoding.}
Six key metrics from -200ms to +500ms at 10ms resolution reveal staged emergence. Red stars indicate peak values. The 78ms resonance lock (vertical red line) corresponds to local minima, confirming it creates simplified coherent substrate. Holographic structure peaks at 170ms (spatial frequency), 230ms (wave coordination), and 270ms (pattern stabilization).}
\label{fig:temporal}
\end{figure}

\textbf{Temporal staging.} High-resolution scanning (10ms steps, -200 to +500ms) revealed staged emergence (Fig~\ref{fig:temporal}): spatial structure peaked at 170ms, wave coordination at 230ms, pattern stabilization at 270ms. The 78ms resonance lock corresponded to local \textit{minima} in spatial complexity, confirming it creates simplified coherent substrate from which complex patterns subsequently emerge.

\begin{figure}[!h]
\centering
\includegraphics[width=\textwidth]{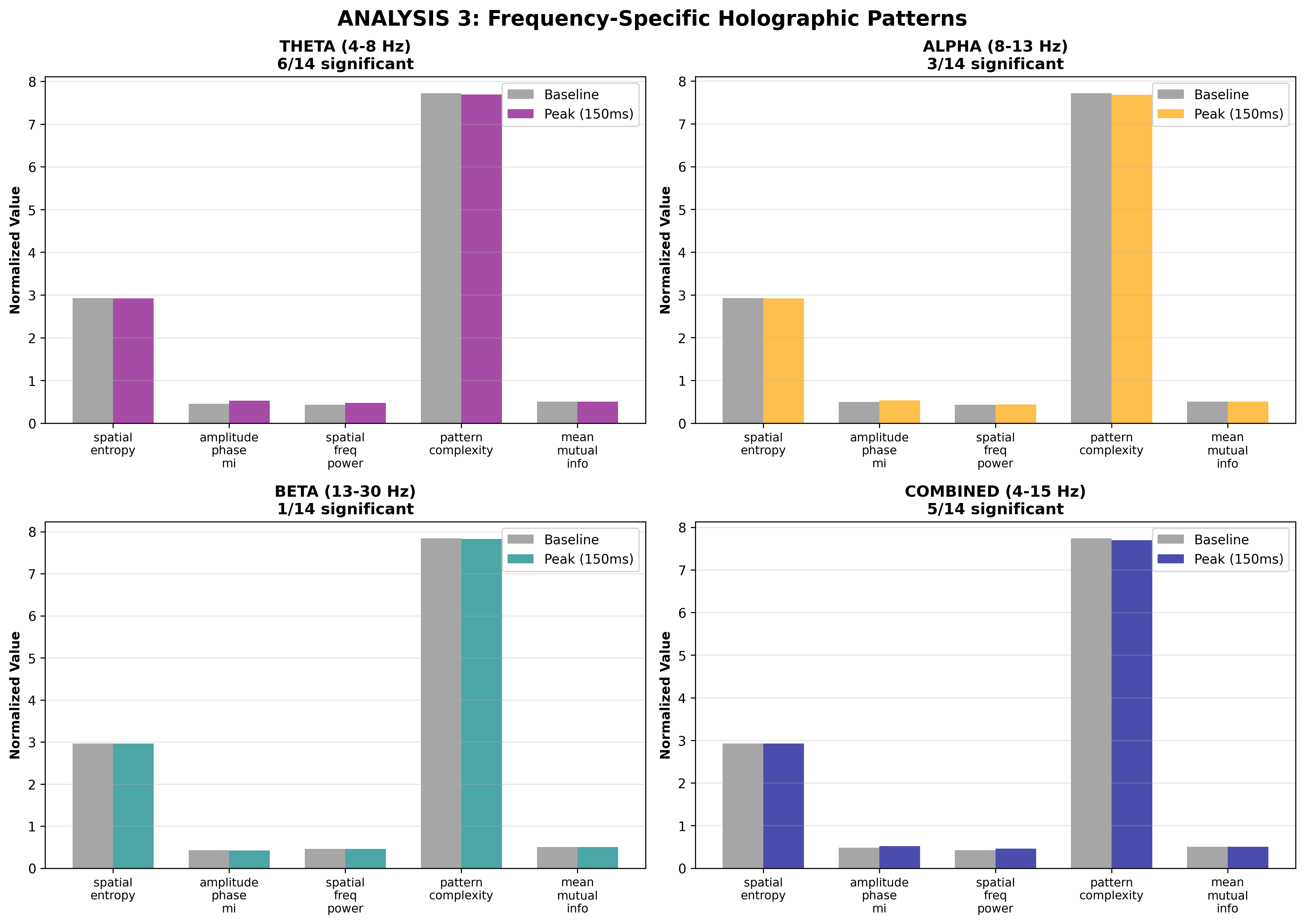}
\caption{\textbf{Theta-band dominance of holographic encoding.}
Normalized metric values at baseline (gray) versus peak 150ms (colored) for each frequency band. Theta (4--8Hz) shows 6/14 significant effects (strongest holographic signature), alpha 3/14, beta 1/14. Progressive weakening at higher frequencies demonstrates frequency-specific encoding consistent with theta's role in long-range coordination and consciousness.}
\label{fig:frequency}
\end{figure}

\textbf{Frequency specificity.} Theta band (4--8Hz) showed 6/14 significant effects versus 3/14 (alpha), 1/14 (beta), indicating holographic encoding is theta-specific (Fig~\ref{fig:frequency}). Theta's long wavelength (125--250ms) enables long-range coordination necessary for distributed encoding\cite{buzsaki2010neuronal}.

\section*{Discussion}

\subsection*{A proposed mechanistic model}

These analyses support a mechanistic model integrating all findings:

\textbf{Stage 1 (continuous): Body Integration.} Physiological artifact removal reduced shared variance between phase synchronization and stimulus-evoked amplitude by approximately 87\%, an effect specific to physiological components (13.8$\times$ specificity ratio). This is consistent with a functional role for physiological signals in maintaining the coupling between large-scale neural coordination and event-related processing.

\textbf{Stage 2 (78ms): Resonance Lock.} Brain and body achieve zero-lag synchronization with simultaneous bidirectional causality, confirmed by partial Granger causality ruling out common drive. This resonance creates coherent substrate analogous to the reference wave in optical holography. The thermodynamic signature---constraint accumulation followed by entropy expansion---indicates a transition from independent oscillators to a coupled system.

\textbf{Stage 3 (78--600ms): Critical Dynamics.} Resonance maintains heavy-tailed avalanche dynamics: Raw data exhibits statistics consistent with near-critical dynamics (power-law or log-normal, heavy-tailed), while Clean data definitively rejects power-law in favor of light-tailed alternatives. Maximum cascade size is 3.6-fold larger with body coupling. The qualitative shift from heavy-tailed to light-tailed distributions is consistent with a transition from near-critical to subcritical regime.

\textbf{Stage 4 (150--270ms): Holographic Encoding.} Complex spatial interference patterns emerge, encoding cognitive state through distributed wave structure. Spatial frequency content increases 18.6\%, amplitude-phase coordination increases 7.6\%, information richness increases 5.5\%.

This proposed sequence addresses several questions: Why does consciousness require time? (Staged construction through physical processes.) Why might embodiment be important? (Provides resonance substrate and critical perturbations.) Why can damage be tolerated? (Distributed holographic encoding.) Why does experience have unified yet differentiated character? (Holographic interference pattern properties.)

\subsection*{Self-organized criticality as mechanism}

Our observed Raw exponent ($\tau = 2.73 \pm 0.15$, 95\% CI [2.63, 3.00]) differs from the classical Bak-Tang-Wiesenfeld sandpile model ($\tau \approx 1.5$)\cite{bak1987self}. Several considerations contextualize this value.

First, critical exponents depend on system dimensionality, conservation laws, and boundary conditions\cite{pruessner2012self}. Neural systems are non-conservative (dissipative synaptic transmission), high-dimensional, and subject to homeostatic regulation---all factors that steepen the size distribution relative to conservative sandpile models. Second, recent theoretical and empirical work demonstrates that SOC exponents in neural systems regularly exceed classical values: exponents above 3 appear in models with inhibitory feedback\cite{poil2012critical} and in empirical neural avalanche studies\cite{priesemann2014spike}. Values approaching 1 are attainable in directed models\cite{bonachela2009self}, demonstrating that the universality class---not a fixed exponent value---defines criticality.

Third, the rigorous goodness-of-fit analysis (see Results) demonstrates that Raw data shows heavy-tailed statistics consistent with power-law or log-normal distributions (indistinguishable from exponential, $p = 0.34$; indistinguishable from log-normal, $p = 0.30$), while Clean data definitively rejects power-law in favor of all tested alternatives (all $p < 0.0001$). This qualitative contrast---heavy-tailed Raw dynamics versus definitively light-tailed Clean dynamics---is consistent with a transition from near-critical to subcritical regime, independent of the exact exponent value.

The preservation of network topology (hierarchical modularity, preferential attachment) despite loss of avalanche dynamics demonstrates dissociation between structure and dynamics. The brain maintains its architectural capacity for criticality, but without body coupling, cannot access critical states. This parallels ferromagnets above Curie temperature: crystal structure persists, but magnetic ordering vanishes.

\subsection*{The coupling argument}

A potential objection claims Raw data simply superimposes random physiological noise onto exponential brain dynamics. This hypothesis fails on multiple grounds:

\textbf{Specificity.} The control analysis demonstrates that removing equivalent non-physiological variance does not replicate the effect (13.8$\times$ specificity ratio). If random variance addition were responsible, any high-variance component removal should produce comparable degradation.

\textbf{Network structure.} Random noise cannot generate preferential attachment ($p < 10^{-10}$). The ``rich-get-richer'' dynamics require systematic coordination impossible for independent random signals.

\textbf{Temporal prediction.} The 78ms resonance \textit{predicts} subsequent P300 amplitude. Random noise cannot predict future states---this requires bidirectional coupling where brain and body mutually constrain each other's dynamics.

\textbf{Information encoding.} Holographic patterns encode stimulus identity and are destroyed by artifact removal. Random superposition cannot encode semantic information; only coupled oscillators generate interference patterns carrying content.

\textbf{Phase relationships.} The Kuramoto parameter measures \textit{phase coherence}, not amplitude. Random phases would decrease $R$; observed high $R$ indicates phase-locking requiring coupling mechanisms.

\textbf{Common drive.} Partial Granger causality retained 100\% of bidirectional coupling after conditioning on global and midline signals, ruling out shared input as the driver of apparent resonance.

\subsection*{Theoretical integration}

Our framework integrates multiple theories while addressing their limitations:

\textbf{Integrated Information Theory}\cite{tononi2016integrated} proposes consciousness corresponds to integrated information $\Phi$. Our holographic framework provides a concrete mechanism: interference patterns naturally satisfy differentiation (distinct phase/amplitude at each location) and integration (mutual information between locations). The distributed encoding principle---each part contains information about the whole---directly implements IIT's integration requirement.

\textbf{Free Energy Principle}\cite{friston2010free} posits cognition minimizes surprise. Our resonance lock corresponds to the moment prediction error is minimized. However, our data suggest that embodiment may not be optional for free energy minimization---humans appear to implement it through constraint-release dynamics dependent on somatic coupling.

\textbf{Enactivist theories}\cite{varela1991embodied} argue cognition is organism-environment interaction. Our data provide quantitative neural-physiological evidence: removing physiological signals reduces phase-voltage coupling to non-significant levels, avalanche dynamics shift from scale-free to exponentially bounded, and holographic encoding metrics are selectively disrupted.

\subsection*{Limitations}

Several important methodological limitations should be acknowledged.

\textbf{Sample size.} This study analyzed 10 subjects with approximately 50 trials per subject. While the within-subject trial-level design provides adequate statistical power for the reported effects, and per-subject analyses demonstrate consistency of the key findings across individuals, the sample size limits generalizability. Replication in larger cohorts is needed to establish population-level reliability.

\textbf{Scalp EEG spatial resolution.} All analyses rely on scalp EEG, which provides only indirect measurements of cortical activity with limited spatial resolution compared to intracranial recordings. The observed phase synchronization patterns may reflect volume-conducted mixtures of deeper sources. While the low imaginary coherence (0.106) and partial Granger causality results argue against volume conduction as the primary driver, intracranial validation using depth electrodes in clinical populations would substantially strengthen these conclusions.

\textbf{ICA separation uncertainty.} Independent Component Analysis provides an imperfect separation of physiological and neural signals. Some ``artifact'' components may contain mixed neural-physiological signals, and some neural components may retain physiological contamination. The control analysis (13.8$\times$ specificity ratio) substantially constrains this concern by demonstrating that the effect is specific to physiological components, but complete certainty about component identity is not achievable with ICA alone.

\textbf{Behavioral correlation.} The P300 matrix speller task required subjects to mentally count target flashes, yielding near-ceiling counting accuracy with insufficient variance to meaningfully correlate with holographic metrics. The task was designed as an attention verification check rather than a graded performance measure. Future studies using tasks that generate graded behavioral performance are needed to establish the behavioral relevance of holographic encoding metrics.

\textbf{Causal interpretation.} While we demonstrate that physiological artifact removal selectively disrupts integrative dynamics, we cannot conclusively establish that physiological signals causally mediate consciousness as opposed to being necessary correlates. The distinction between mediation and correlation requires interventional studies beyond the scope of EEG analysis.

\subsection*{Future perspectives}

The following theoretical extrapolations are motivated by our empirical findings but extend beyond what the present data can directly demonstrate.

\textbf{Implications for artificial intelligence.} Current AI systems excel at pattern recognition but face challenges in causal reasoning\cite{pearl2018book}. Our findings suggest that this may partly reflect architectural constraints: if conscious integration requires the specific thermodynamic properties we observe---embodied constraints, stochastic perturbations, resonant coupling, and thermodynamic costs---then deterministic digital architectures may face fundamental limitations. Neuromorphic approaches implementing analog, stochastic, embodied systems with intrinsic dynamics\cite{indiveri2011neuromorphic} may be required to achieve comparable integrative capacity. These predictions remain theoretical and await empirical testing.

\textbf{Clinical applications.} If consciousness requires specific thermodynamic architecture, disorders of consciousness may reflect disrupted brain-body coupling. Measuring avalanche exponents, resonance strength, and holographic metrics could provide objective markers for consciousness levels in coma, vegetative state, or minimally conscious patients. This possibility warrants investigation in clinical populations.

\textbf{The hard problem.} Why should physical processes give rise to subjective experience\cite{chalmers1995facing}? Our framework suggests a possible response: if certain physical processes---interference patterns in coupled oscillatory systems at criticality---\textit{constitute} conscious experiences rather than merely correlating with them, the hard problem may be reframed. This remains a philosophical position informed by, but not proven by, the present data.

\section*{Conclusion}

We have presented evidence that human cognitive integration involves self-organized criticality maintained by brain-body resonance. Removal of physiological signals reduces shared variance between phase synchronization and stimulus-evoked amplitude by approximately 87\%, an effect specific to physiological components (13.8$\times$ specificity ratio). Brain-body resonance at 78ms creates zero-lag synchronization with simultaneous bidirectional causality, confirmed by partial Granger causality analysis ruling out common drive. Raw data shows heavy-tailed avalanche dynamics ($\tau = 2.73 \pm 0.15$) consistent with near-critical dynamics, while Clean data definitively rejects power-law in favor of all tested alternatives (all $p < 0.0001$)---a qualitative contrast consistent with a transition from near-critical to subcritical regime. Critical dynamics are associated with holographic information encoding with spatial frequency power increasing 18.6\% ($p < 0.0001$) between 150--270ms post-resonance.

These findings challenge the assumption that cognition can be fully characterized by isolated neural computation, and suggest that physiological signals selectively support the integrative dynamics indexed by phase synchronization. The dissociation between preserved local processing (P300) and disrupted global integration (phase coherence, avalanche statistics, holographic encoding) points to a distinction between local cortical computation and whole-body integrative dynamics that warrants further investigation across larger samples, clinical populations, and complementary neuroimaging modalities.

\section*{Methods}

\subsection*{Dataset and participants}

\textbf{Dataset.} Publicly available 64-channel EEG from PhysioNet ERP-BCI dataset\cite{citi2010documenting}. Ten subjects (5 male, 5 female, ages 23--51) performed visual oddball task. Total 500 target trials after quality control. Recording: 2048Hz BioSemi ActiveTwo, downsampled to 256Hz for analysis.

\textbf{Ethics.} This study analyzed publicly available de-identified data. Original data collection received appropriate ethical approval as documented in ref. \cite{citi2010documenting}. No new human subjects research was conducted for this study.

\subsection*{Preprocessing pipelines}

Two parallel pipelines were established to test the artifact hypothesis:

\textbf{Raw (whole-body):} (1) Average reference, (2) 4--30Hz bandpass (zero-phase FIR), (3) Epochs -500ms to +3500ms relative to stimulus onset. This preserves all physiological signals.

\textbf{Clean (artifact-rejected):} Identical to Raw plus Independent Component Analysis (ICA) artifact rejection. EOG components (2--3 per subject) were identified via correlation with frontal electrodes ($r > 0.7$) and removed. This implements conventional preprocessing.

\textbf{Control conditions:} To test specificity (see Control Analysis in Results), two additional conditions were computed from the same ICA decomposition: (1) removal of the same number of highest-variance non-physiological components, and (2) removal of randomly selected non-physiological components (bootstrapped over 20 iterations per subject).

\subsection*{Phase synchronization analysis}

\textbf{Kuramoto order parameter.} Global phase coherence quantified as:
\begin{equation}
R(t) = \left| \frac{1}{N} \sum_{j=1}^{N} e^{i\phi_j(t)} \right|
\end{equation}
where $\phi_j(t)$ is instantaneous phase of channel $j$ via Hilbert transform, $N$ is number of channels. Phase synchronization is an established neural correlate of large-scale cognitive integration\cite{varela2001brainweb,lachaux1999measuring}.

\textbf{Phase Slope Index.} Directional coupling between posterior (Pz, PO3, PO4, PO7, PO8) and frontal (Fp1, Fp2, AF3, AF4, F7, F8) regions computed following ref. \cite{nolte2008robustly} in 4--15Hz band. PSI $\approx 0$ with high coherence indicates zero-lag synchronization.

\subsection*{Granger causality}

Time-lagged ridge-regularized Granger causality analysis tested bidirectional coupling between brain and body regions. For each lag $\tau \in [0, 600]$ ms (step: 10ms), we tested whether past activity in region $X$ predicts future activity in region $Y$ beyond $Y$'s autoregressive history using model order $p = 3$ and regularization $\alpha = 0.01$ (chosen via cross-validation). F-statistics quantified prediction improvement.

\textbf{Partial Granger causality.} To rule out common drive artifacts, we repeated the analysis conditioning on (1) the global mean signal across all electrodes, and (2) midline electrode signals (Cz, Pz, Fz), comparing peak F-statistics and retention ratios across conditions.

\subsection*{Thermodynamic analysis}

\textbf{Sample entropy.} Signal regularity quantified using sliding 200ms windows (25ms steps) following ref. \cite{richman2000physiological} with parameters $m = 2$ (embedding dimension), $r = 0.2 \times$ SD (tolerance).

\textbf{Phase space construction.} Two-dimensional phase portraits plotted causal drive (peak Granger F-statistic) versus state entropy at each timepoint, revealing thermodynamic trajectories analogous to pressure-volume diagrams.

\subsection*{Criticality tests}

\textbf{Avalanche detection.} Continuous periods where $R(t)$ exceeded 75th percentile threshold. Size and duration distributions analyzed using the Clauset-Shalizi-Newman (2009) method\cite{clauset2009power}: optimal $x_{\min}$ estimated via KS distance minimization, exponents fitted by maximum likelihood, and likelihood ratio tests compared power-law against exponential, log-normal, stretched exponential, and truncated power-law alternatives. Bootstrap confidence intervals on exponents computed over 1,000 iterations. Implemented using the \texttt{powerlaw} Python package\cite{alstott2014powerlaw}.

\textbf{Spatial organization.} Phase-locking values between all electrode pairs: $\text{PLV}_{jk} = |\langle e^{i(\phi_j - \phi_k)} \rangle_{\text{trials}}|$. Distance-correlation analysis identified hierarchical modularity (ratio of global/local length scales).

\textbf{Network topology.} Dynamic functional networks constructed in 50ms sliding windows, edges defined by 90th percentile PLV threshold. Degree distributions and preferential attachment tested across 73 temporal networks.

\subsection*{Holographic analysis}

Fourteen spatial metrics computed spanning: (1) Information theory (spatial entropy, amplitude entropy, phase variance, pattern complexity), (2) Wave interference (spatial correlation, amplitude-phase mutual information), (3) Spatial frequency (2D FFT power, dominant frequency, spectral centroid on interpolated 32$\times$32 grids), (4) Distributed encoding (mean/max/variance of pairwise mutual information), (5) Reproducibility (within-trial and cross-trial pattern similarity). Metrics computed at 150ms (peak) versus -200ms (baseline) in theta (4--8Hz), alpha (8--13Hz), beta (13--30Hz), and combined (4--15Hz) bands.

\subsection*{Statistical analysis}

Independent samples $t$-tests compared peak versus baseline states. Cohen's $d$ quantified effect sizes. Pearson correlations tested relationships between variables. One-sample $t$-tests evaluated PSI against zero. Likelihood ratio tests compared distribution models. Significance threshold: $\alpha = 0.05$, two-tailed. All analyses performed trial-by-trial, then session-averaged, then subject-aggregated. Results reported as mean $\pm$ SEM unless otherwise specified. Per-subject analyses are reported for all key findings to assess consistency across the sample.

\subsection*{Software and code availability}

All analyses implemented in Python 3.9 using MNE-Python 1.0\cite{gramfort2013meg}, SciPy 1.9, NumPy 1.23, Matplotlib 3.6, and the \texttt{powerlaw} package\cite{alstott2014powerlaw}. All analysis code, including control analyses, is publicly available at \url{https://github.com/wadamalon/frankenstein}.

\section*{Data availability}

Source data are publicly available from the PhysioNet repository at https://doi.org/10.13026/C2101S. All analysis code, including control analyses added during revision, is publicly available at https://github.com/wadamalon/frankenstein. Processed intermediate data supporting the reported statistics are provided as Supporting Information files.

\section*{Acknowledgments}

The author thanks Professor Ehab Emam for invaluable guidance throughout this project. The author acknowledges Claude (Anthropic) for assistance with manuscript preparation, figure organization, and LaTeX formatting. Source data were obtained from PhysioNet\cite{citi2010documenting}.

\section*{Author contributions}

A.G.E.: Conceptualization, Methodology, Software, Formal Analysis, Investigation, Writing (original draft and editing), Visualization.

\section*{Competing interests}

The author declares no competing interests.

\section*{Supporting information}

\paragraph*{S1 Fig.}
\label{S1_Fig}
\textbf{Per-subject results for all key analyses.} Individual subject data for phase-voltage coupling, brain-body resonance, avalanche dynamics, and holographic encoding metrics.

\paragraph*{S1 Table.}
\label{S1_Table}
\textbf{Per-subject summary statistics.} Complete statistical results for each of the 10 subjects across all four analyses.

\paragraph*{S2 Table.}
\label{S2_Table}
\textbf{Trial-level R-ERP correlation data.} Processed intermediate data supporting the phase-voltage coupling analysis.

\paragraph*{S3 Table.}
\label{S3_Table}
\textbf{Avalanche size distribution data.} Processed intermediate data supporting the criticality analysis.

\paragraph*{S4 Table.}
\label{S4_Table}
\textbf{Holographic encoding metrics.} Processed intermediate data supporting the holographic analysis.

\paragraph*{S5 Table.}
\label{S5_Table}
\textbf{Brain-body resonance parameters.} Granger causality F-statistics and Phase Slope Index values across trials.

\paragraph*{S6 Table.}
\label{S6_Table}
\textbf{Avalanche summary statistics.} Complete avalanche summary data including size distributions for Raw and Clean conditions.

\nolinenumbers


\end{document}